\documentclass[submission,copyright,creativecommons]{eptcs}
\DeclareMathAlphabet{\mathcal}{OMS}{cmsy}{m}{n}  % To get the right mathcal font
\providecommand{\event}{LSFA 2021} % Name of the event you are submitting to
\usepackage{underscore}           % Only needed if you use pdflatex.
\hypersetup{
  colorlinks=true,
  linkcolor=blue,
  anchorcolor=blue,
  citecolor=blue,
  urlcolor=black,
  filecolor=blue,
  breaklinks=true
}

\usepackage{cite}
\usepackage{amsmath}
\usepackage{amsthm}
\usepackage{amssymb}
\usepackage{amsfonts}
\usepackage{proof}
\usepackage[T1]{fontenc}
\usepackage{stmaryrd}

\newtheorem{theorem}{Theorem}[section]
\theoremstyle{definition}
\newtheorem{remark}[theorem]{Remark}
\newtheorem{definition}[theorem]{Definition}
\newtheorem{example}[theorem]{Example}
\newtheorem{postulate}{Postulate}

\newcommand\ket[1]{\ensuremath{|{#1}\rangle}}
\newcommand\bra[1]{\ensuremath{\langle{#1}|}}
\newcommand\braket[2]{\ensuremath{\langle{#1}|{#2}\rangle}}
\newcommand\lra{\ensuremath{\longrightarrow}}
\newcommand\sem[1]{\llbracket #1\rrbracket}
\newcommand\irule[3]{\infer[{\mbox{\footnotesize $#3$}}]{#2}{#1}}
\newcommand\pair[2]{\langle #1, #2\rangle}

\title{A Quick Overview on the\\ Quantum Control Approach to the Lambda Calculus}
\author{
  Alejandro D\'iaz-Caro\thanks{Founded by PICT-2019-1272, STIC-AmSud 21STIC10 Qapla', ECOS-Sud A17C03 QuCa, and PIP 11220200100368CO.}
  \institute{
      \begin{tabular}{c@{\quad}c}
	\parbox[t]{6cm}{\centering
	  Departamento de Ciencia y Tecnolog\'ia\\
	  Universidad Nacional de Quilmes\\
	  Bernal, Buenos Aires, Argentina
	}
	&
	\parbox[t]{6cm}{\centering
	    Instituto de Ciencias de la Computaci\'on\\
	    CONICET--Universidad de Buenos Aires\\ 
	    Buenos Aires, Argentina
	}
      \end{tabular}
  }
  \email{adiazcaro@icc.fcen.uba.ar}
}
\def\titlerunning{A Quick Overview on the Quantum Control Approach to the Lambda Calculus}
\def\authorrunning{A. D\'iaz-Caro}
\begin{document}
\maketitle

\begin{abstract}
  In this short overview we start with the basics of quantum computing,
  explaining the difference between the quantum and the classical control
  paradigms. We give an overview of the quantum control line of research within
  the lambda calculus, ranging from untyped calculi up to categorical and
  realisability models. This is a summary of the last 10+ years of research in
  this area, starting from Arrighi and Dowek's seminal work until today.
\end{abstract}

\section{Introduction}
The study of quantum computing in the framework of the lambda calculus has more than one motivation. 

On the one hand, it is a tool to develop programming languages with firm 
foundations. For example, from the study of Selinger and Valiron's ``Quantum
Lambda Calculus'' (QLC)~\cite{SelingerValironMSCS06},
Quipper~\cite{GreenLeFanulumsdaineRossSelingerValironPLDI13} and
QWIRE~\cite{Qwire} arose. Both quantum languages are quite advanced and
complex, and while they are not fully formalised, their cores are based on
Knill's QRAM model~\cite{KnillTR96}, which proposes that a quantum computer is
a device attached to a classical computer, and it is the classical computer
which instructs the quantum computer on what operations to perform, over which
qubits, etc. Selinger took this QRAM model and formalised it in what is called
the ``Quantum Data, Classical Control'' approach~\cite{SelingerMSCS04}. This
approach derived later into the QLC, and finally in Quipper and QWIRE. Most
quantum programming languages follow the same model, including high level
languages such as~IBM's Qiskit~\cite{Qiskit} or Microsoft's
Q$\sharp$~\cite{Qsharp}.  Indeed, the Quantum Data, Classical Control approach
is the more practical approach.

On the other hand, typed lambda calculus provides a way to study logics,
through the Curry-Howard isomorphism~(see, for
example,~\cite{SorensenUrzyczyn98}), which connects logics with computation.
The logic of quantum mechanics has been a challenging subject since the
beginnings of quantum mechanics. The seminal work of Birkhoff and von
Neumann~\cite{BirkhoffvonNeumann36}, in which the authors tried to reconcile
the apparent inconsistency of classical logic with the quantum measurement, was
the birth of a long field of study among physicists. However, due to its
origins, there is no formal connection of this line with computing, and so, an
extension to the Curry-Howard isomorphism contemplating this logic is not easy
to envisage. The apparition of quantum computing as a way of understanding
quantum mechanics brought all the computer science machinery to the game. As a
consequence, a quantum logic formally connected to a quantum lambda calculus
seems reasonable.  If Birkhoff and von Neumann's quantum logic takes into
account the superposition of quantum states, and the uncertainty principle, a
quantum lambda calculus aimed at being the basis for a quantum logic must also
consider these non-classical aspects of the quantum theory. With this objective
in mind is that the first developments in this direction dropped the idea of Quantum
Data, Classical Control, embracing Quantum Control instead. 

The quantum control approach has several origins. Most notably, quantum
automata exploit quantum control.  The main difference with classical control
is that in classical control the control flow of a program is classical in the
sense that its description does not admit superpositions, measurement, or any
other kind of quantum properties. For example, we can describe a quantum
algorithm by describing its classical flow such as: 

\texttt{Prepare a quantum state $\ket{\psi}$}

\texttt{Apply the unitary transformation $U$ to $\ket{\psi}$}

\texttt{Apply the unitary transformation $V$ to the result of the previous step}

\texttt{Measure the obtained system}

\noindent There is nothing quantum in the flow described here. This list of
instructions can be done by a classical computer. The quantum computer is the
device which will have to perform the operations prescribed by this classical
machine. In quantum control instead the flow is not classical. For example, in
the instruction
\[
  \mathtt{If}\ c\ \mathtt{then}\ \ket{\psi}\ \mathtt{else}\ \ket{\varphi}
\]
if we admit $c$ to be a superposition such as
$\alpha.\ket{\mathsf{true}}+\beta.\ket{\mathsf{false}}$, this instruction would
become the superposition
\[
  \alpha.\bigg(\mathtt{If}\ \ket{\mathsf{true}}\ \mathtt{then}\ \ket{\psi}\ \mathtt{else}\ \ket{\varphi}\bigg)
  \ +\ 
  \beta.\bigg(\mathtt{If}\ \ket{\mathsf{false}}\ \mathtt{then}\ \ket{\psi}\ \mathtt{else}\ \ket{\varphi}\bigg)
\]
and so resulting in
\[
  \alpha.\ket{\psi}+\beta.\ket{\varphi}
\]
The control flow is in superposition, and so it is not classic. Of course this
instruction is only valid if the final state
$\alpha.\ket{\psi}+\beta.\ket{\varphi}$ is a valid quantum state, which is not
always the case.  In particular, if the norm of the input
$\alpha.\ket{\mathsf{true}}+\beta.\ket{\mathsf{false}}$ being $1$ implies that
the norm of the output is also $1$, then this ``quantum if'' instruction can be
implemented by a quantum operator. The first work in this line, defining the
quantum-if in the lambda calculus, was that of Altenkirch and
Grattage~\cite{AltenkirchGrattageLICS05}.  Since then, there has been a
long line of independent research pursuing a ``quantum computational logic'',
that is, some sort of quantum logic firmly founded on quantum computing.

This article intends to be an overview on this quest for a quantum
computational logic.

\paragraph*{Plan of the paper}
In Section~\ref{sec:QC} we give a brief introduction to the
quantum computing formalism.
In Section~\ref{sec:Lineal} we present Lineal, an untyped extension to the
lambda calculus to deal with superpositions. It is the starting point on the
quest for a quantum computational logic.
In Section~\ref{sec:Vectorial} we present Vectorial and some of its fragments,
which is the first typed Lineal.
In Section~\ref{sec:LambdaS} we present Lambda-$\mathcal S$, which is another
typed Lineal, extended with quantum measurements. We also provide a categorical
interpretation of this calculus.
In Section~\ref{sec:LambdaSone} we show the first restriction of Lineal into a
quantum language, achieved by using realisability techniques. We then introduce
the calculus Lambda-$\mathcal S_1$, which has been derived from this technique,
and give some details on its categorical interpretation. Such an interpretation
is  related to that of Lambda-$\mathcal S$.
In Section~\ref{sec:Sup} we move from the opposite direction: we define a new
connective $\odot$ (read ``sup'') in Natural Deduction, which induces the
$\odot$-calculus. Then, we show how to transparently add complex scalars,
defining the $\odot^{\mathbb C}$-calculus, and use it to encode a quantum
language.
In Section~\ref{sec:recursive} we refer to a few recent works towards recursive
types in the quantum control setting.
Finally, we conclude in Section~\ref{sec:Conclusion} with some final thoughts
and open problems.

\section{Quantum computing in terms of four postulates}
\label{sec:QC}
This section does not pretend to be an extensive introduction to quantum
computing but just for the basics, and we take the liberty to simplify many
things for the sake of readability. For a quite complete introduction to
quantum computing the reader is referred to the great book by Nielsen and
Chuang~\cite{NielsenChuang00}. 

Quantum mechanics can be described in four axioms or postulates. 

The first postulate defines how the quantum states are represented.
\begin{postulate}[State space]\label{pos:one}
  The state of an isolated quantum system can be fully described by a {\em
  state vector}, which is a norm-$1$ vector in a Hilbert space, that is, Banach
  space with inner product.
\end{postulate}
In quantum computing we usually consider the Hilbert space $\mathbb C^{2^n}$,
hence, from now on we only consider these spaces.
For vectors in $\mathbb C^{2^n}$ we use the Dirac notation consisting of a
binary encoding on vectors.  For example, 
\begin{align*}
  \left(\begin{smallmatrix}
      \frac{\sqrt 3}2\\
      \frac 12 
  \end{smallmatrix}\right)\in\mathbb C^2
  &
  \mbox{ is written $\frac{\sqrt 3}2\ket{0}+\frac 12\ket{1}$,}
  \\
  \mbox{and}\quad
  \left(\begin{smallmatrix} 
      \frac 1{\sqrt 3}\\
      0\\ 
      \frac 1{\sqrt 3} \\
      \frac 1{\sqrt 3} 
  \end{smallmatrix}\right)\in\mathbb C^4
  &
  \mbox{ is written $\frac 1{\sqrt 3}\ket{00}+\frac 1{\sqrt 3}\ket{10}+\frac 1{\sqrt 3}\ket{11}$.}
\end{align*}
A generic state vector is written $\ket\psi$. We also write $\bra\psi$ to the
transpose conjugate of $\ket\psi$. This way, $\ket\psi\bra\varphi$ is a matrix
while $\bra\psi\ket\varphi$ (usually written $\braket\psi\varphi$) is the inner
product $(\ket\psi,\ket\varphi)$.

The second postulate defines how a quantum state evolves over time. We give its
discrete time version, since in quantum computing we usually consider discrete
time.
\begin{postulate}[Evolution]\label{pos:two}
  The {\em evolution} of an isolated quantum system can be described by a
  unitary matrix, that is, a matrix $U$ such that $U^\dagger = U^{-1}$, where
  $U^\dagger$ is the conjugate transpose of $U$. If the state of a quantum
  system is described by $\ket\psi$, after the evolution $U$ the new state is
  $\ket\varphi = U\ket\psi$.
\end{postulate}

\begin{example}\label{ex:H}
  Let $H = \left(
    \begin{smallmatrix}
      \frac 1{\sqrt 2} & \frac 1{\sqrt 2}\\
      \frac 1{\sqrt 2} & \frac {-1}{\sqrt 2}
    \end{smallmatrix}
  \right)$, then 
  \begin{align*}
    H\ket 0 &= \frac 1{\sqrt 2}\ket 0+\frac 1{\sqrt 2}\ket 1 = \ket+\\
    H\ket 1 &= \frac 1{\sqrt 2}\ket 0+\frac {-1}{\sqrt 2}\ket 1 = \ket-
  \end{align*}
  where $\ket+$ and $\ket-$ are just conventional notations for these vectors.
  In particular, notice that $\{\ket+,\ket -\}$ is an orthonormal basis of
  $\mathbb C^2$ as well as $\{\ket 0,\ket 1\}$. Hence, $H$ is a basis change
  matrix.

  $H$ is known as the Hadamard gate.
\end{example}

The third postulate defines how a quantum system is measured. We give its
general form as Postulate~\ref{pos:three}, and a way to simplify it in
Theorem~\ref{thm:meas}.

\begin{postulate}[Quantum measurement]\label{pos:three}
  The {\em quantum measurement} is described by a collection of square matrices
  $\{M_i\}_i$, where $i$ is called the {\em output of the measurement}, such that
  \[
    \sum_i M_i^\dagger M_i = I
  \]
  If the state of a quantum system is described by $\ket\psi$, the probability
  of measuring $i$ is given by $p_i = \bra\psi M_i^\dagger M_i\ket\psi$ and the
  state after the measuring $i$ is $\ket\varphi = \frac{1}{\sqrt{p_i}}
  M_i\ket\psi$.
\end{postulate}
\begin{example}
  Consider the following measurement: $\{M_0,M_1\}$, with $M_0 = \ket 0\bra 0$
  and $M_1 = \ket 1\bra 1$. That is, 
  $M_0 = \left(
    \begin{smallmatrix}
      1 & 0\\
      0 & 0
    \end{smallmatrix}
  \right)$
  and
  $M_1 = \left(
    \begin{smallmatrix}
      0 & 0\\
      0 & 1
    \end{smallmatrix}
  \right)$, and let $\ket\psi = \frac{\sqrt 3}2\ket 0+\frac 12\ket 1$.

  Then, $p_0 =\braket\psi 0\braket 00\braket 0\psi = \frac 34$ and $p_1 =
  \braket\psi 1\braket 11\braket 1\psi =\frac 14$. The state after measuring
  $0$ is $\ket 0$ and after measuring $1$ is $\ket 1$.
\end{example}
The previous example can be taken as a general rule.  With these $M_0$ and
$M_1$, in general, measuring $\alpha\ket 0+\beta\ket 1$ results in $\ket 0$
with probability $|\alpha|^2$ and $\ket 1$ with probability $|\beta|^2$.

Moreover, taking $\{P_b\}_{b\in \{0,1\}^n}$ with $P_b = \ket b\bra b$ to
measure the state $\ket\psi = \sum_b\alpha_b\ket b$ results in $\ket b$ with
probability $|\alpha_b|^2$.

The following theorem states that such a measurement (usually called
``measurement in the computational basis'') is enough for quantum computing.
\begin{theorem}
  \label{thm:meas}
  Any measurement $M = \{M_i\}_i$ can be simulated by a measurement in the
  computational basis given by $P=\{P_b\}_{b\in \{0,1\}^n}$ with $P_b = \ket
  b\bra b$, and a unitary matrix $U$ in the sense that measuring $\ket\psi$
  with $M$ is the same as measuring $U\ket\psi$ with $P_b$ and applying
  $U^{-1}$ afterwards.
  \qed
\end{theorem}

The previous lemma justifies the fact that most quantum programming languages
consider only measurements in the computational basis.

Finally, the fourth postulate defines how to compose quantum systems.

\begin{postulate}[Composed system]\label{pos:four}
  The state space of a {\em composed system} is the tensor product of the state
  of its components.

  Given $n$ systems in states $\ket{\psi_1},\dots,\ket{\psi_n}$, the composed
  system is described by 
  \(
    \ket{\psi_1}\otimes\cdots\otimes\ket{\psi_n}
  \).
\end{postulate}

\begin{example}
  If $\ket\psi = \frac 1{\sqrt 2}\ket 0+\frac 1{\sqrt 2}\ket 1$ and
  $\ket\varphi = \frac 1{\sqrt 5}\ket 0+\frac 2{\sqrt 5}\ket 1$, the composed
  system $\ket\psi\otimes\ket\varphi$ is
  \[
    \left(\frac 1{\sqrt 2}\ket 0+\frac 1{\sqrt 2}\ket 1\right)
    \otimes
    \left(\frac 1{\sqrt 5}\ket 0+\frac 2{\sqrt 5}\ket 1\right)
    =
    \frac 1{\sqrt{10}}\ket{00} +
    \frac 2{\sqrt{10}}\ket{01} +
    \frac 1{\sqrt{10}}\ket{10} +
    \frac 2{\sqrt{10}}\ket{11}
  \]
\end{example}

\begin{definition}[qubit]
  We call qubit, or quantum bit, to the quantum states in $\mathbb C^2$ and
  $n$-qubit to quantum states in $\mathbb C^{2^n} = \mathbb
  C^2\otimes\cdots\otimes\mathbb C^2$.

  So, a qubit is written $\alpha\ket 0+\beta\ket 1$ and an $n$-qubit is
  $\sum_{b\in\{0,1\}^n}\alpha_b\ket b$.
\end{definition}

An important surprisingly consequence of the four postulates is the no-cloning
theorem~\cite{WoottersZurekNATURE82}, which states that an unknown quantum
state cannot be duplicated.
\begin{theorem}
  \label{thm:no-cloning}
  There is no unitary matrix $U$ such that for some fixed
  $\ket\varphi\in\mathbb C^{2^n}$ and for all $\ket\psi\in\mathbb C^{2^n}$ we
  have $U(\ket\varphi\otimes\ket\psi)=\ket\psi\otimes\ket\psi$.
  \qed
\end{theorem}

\section{Lineal: A linear algebraic lambda calculus}
\label{sec:Lineal}
Lineal, a seminal work by Arrighi and Dowek, first published at
\cite{ArrighiDowekRTA08} and then extended in \cite{ArrighiDowekLMCS17}, starts
from the following simple idea: The Church encoding for booleans, where
$\lambda x.\lambda y.x$ represents $\mathsf{true}$ and $\lambda x.\lambda y.y$
represents $\mathsf{false}$, follows the premise that in the lambda calculus
everything is a function, even the basic data. Therefore, to consider quantum
bits we would need to extend the lambda calculus with complex linear
combinations of lambda terms. This way, the qubit $\alpha\ket 0+\beta\ket 1$
would be represented by $\alpha.\lambda x.\lambda y.x + \beta.\lambda x.\lambda
y.y$. More generally, we may consider the infinite-dimensional Hilbert space
whose basis is given by the values of the lambda calculus.

Hence, Lineal proposes the following syntax of terms:
\[
  t := x\mid \lambda x.t\mid tt\mid \alpha.t\mid t+t\mid \vec 0
\]
where $\alpha\in\mathbb C$ and the symbol $+$ is considered modulo associativity and commutativity.

Its operational semantics has four groups of rules.

The beta group:
\[
  (\lambda x.t)b \lra (b/x)t
\]
where $b$ is a basis term, that is, a classical value (either a variable or an abstraction).

The elementary group:
\[
  \begin{array}{r@{\ }l@{\qquad\qquad}r@{\ }l@{\qquad\qquad}r@{\ }l}
    t+\vec 0&\lra t
    &
    0.t&\lra \vec 0
    &
    1.t&\lra t\\
    \alpha.\vec 0&\lra\vec 0
    &
    \alpha.(\beta.t)&\lra(\alpha\times\beta).t
    &
    \alpha.(t+r)&\lra\alpha.t+\beta.t  
  \end{array}
\]

The factorization group:
\[
  \begin{array}{r@{\ }l@{\qquad\qquad}r@{\ }l@{\qquad\qquad}r@{\ }l}
    \alpha.t+\beta.t&\lra(\alpha+\beta).t
    &
    \alpha.t+t &\lra(\alpha+1).t
    &
    t+t&\lra 2.t
  \end{array}
\]

The application group:
\[
  \begin{array}{r@{\ }l@{\qquad\qquad}r@{\ }l@{\qquad\qquad}r@{\ }l}
    (t+r)s &\lra (ts)+(rs)
    &
    (\alpha.t)r &\lra \alpha.(tr)
    &
    \vec 0 t &\lra \vec 0
    \\
    s (t+r) &\lra (st)+(sr)
    &
    r(\alpha.t) &\lra\alpha.(rt)
    &
    t\vec  0&\lra\vec 0
  \end{array}
\]

This way, for example, if $\ket 0 = \lambda x.\lambda y.x$ and $\ket 1 =
\lambda x.\lambda y.y$, a term encoding a unitary matrix $U$ will act as
follows:
\[
  U(\alpha.\ket 0+\beta.\ket 1) \lra (U\alpha.\ket 0)+(U\beta.\ket 1)\lra^*\alpha.U\ket 0+\beta.U\ket 1
\]

For instance, the Hadamard gate (see Example~\ref{ex:H}) can be encoded as
\begin{equation}
  \label{eq:H}
  H := \lambda x.\left\{x \left[\frac 1{\sqrt 2}.\ket 0 + \frac 1{\sqrt 2}.\ket 1\right] \left[\frac 1{\sqrt 2}.\ket 0 + \frac {-1}{\sqrt 2}.\ket 1\right]\right\}
\end{equation}
where $[t] := \lambda x.t$ is a thunk and $\{t\} := t\lambda x.x$ releases the
thunk (i.e.~$\{[t]\}\lra^* t$).

The thunk is used to stop the linearity. Otherwise, $H\ket 0\lra\ket 0 \left(\frac 1{\sqrt
2}.\ket 0 + \frac 1{\sqrt 2}.\ket 1\right)\left(\frac 1{\sqrt 2}.\ket 0 +
\frac{-1}{\sqrt 2}.\ket 1\right)$, which is just $(\lambda x.\lambda
y.x)\left(\frac 1{\sqrt 2}.\ket 0 + \frac 1{\sqrt 2}.\ket 1\right)\left(\frac
1{\sqrt 2}.\ket 0 + \frac{-1}{\sqrt 2}.\ket 1\right)$ would reduce, using the
rules at the application group, to 
\[
  \frac 12.(\lambda x.\lambda y.x)\ket 0\ket 0
  +
  \frac {-1}2.(\lambda x.\lambda y.x)\ket 0\ket 1
  +
  \frac 12.(\lambda x.\lambda y.x)\ket 1\ket 0
  +
  \frac {-1}2.(\lambda x.\lambda y.x)\ket 1\ket 1
\]
and then to
\[
  \frac 12.\ket 0
  +
  \frac {-1}2.\ket 0
  +
  \frac 12.\ket 1
  +
  \frac {-1}2.\ket 1
  \lra^*
  \vec 0
\]
instead of the expected $\frac 1{\sqrt 2}.\ket 0 + \frac 1{\sqrt 2}.\ket 1$.

So, the linearity achieved by the rules at the application group can be frozen
with thunks when needed, making Lineal a very expressive calculus of
computation combining matrices and vectors.

There is no measurement in Lineal. In addition, the rules of the operational
semantics have some constraints such as 
\begin{equation}
  \label{eq:restriction}
  \mbox{``rule $\alpha.t+\beta.t\lra(\alpha+\beta).t$ applies only if $t$ is
  closed normal''}
\end{equation}
These constraints are needed in this untyped case where non-terminating terms
can break confluence. For example, let $\Delta_b = \lambda x.(xx+b)$ and $Y_b =
\Delta_b\Delta_b$. Then $Y_b\lra Y_b+b$. If $b$ represents $1$, this $Y_b$ can
add up to infinity. Then, since we have infinity and algebraic operations we
may run into undefined forms such as $\infty-\infty$, which here is represented
by $Y_b + (-1).Y_b$. Indeed, without any restrictions, $Y_b + (-1).Y_b \lra
0.Y_b\lra \vec 0$, but also $Y_b + (-1).Y_b \lra Y_b + b + (-1).Y_b\lra^*b$.
Hence, restriction \eqref{eq:restriction} removes this form of indeterminacy.
There are other restrictions, but since we will consider types in the next
section, which ensures strong normalisation, we can just ignore them.

\begin{remark}
  \label{rmk:no-cloningLineal}
  The algebraic linearity of Lineal implies that the no-cloning theorem
  (Theorem~\ref{thm:no-cloning}) is valid using this encoding of unitary
  matrices. Intuitively, the fact that $U(\alpha.\ket 0+\beta.\ket
  1)\lra^*\alpha.U\ket 0+\beta.U\ket 1$ means that $\alpha$ and $\beta$ are
  never duplicated by $U$, and so it is not possible to construct a term $U$
  such that $U(\alpha.\ket 0+\beta.\ket 1)$ reduce to $(\alpha.\ket
  0+\beta.\ket 1)\otimes(\alpha.\ket 0+\beta.\ket 1)$, for any encoding of
  $\otimes$.
\end{remark}

\section{Vectorial: The first typed Lineal}
\label{sec:Vectorial}
Typing Lineal with simply types, or second order polymorphic types, is possible
by adding the following straightforward typing rules to simply typed lambda
calculus or System F, in order to type the extra constructions:
\[
  \infer{\Gamma\vdash\vec 0:A}{}
  \qquad\qquad
  \infer{\Gamma\vdash t+r:A}{\Gamma\vdash t:A & \Gamma\vdash r:A}
  \qquad\qquad
  \infer{\Gamma\vdash\alpha.t:A}{\Gamma\vdash t:A}
\]
This way, the term $H$ from Equation~\eqref{eq:H} would be typed with
$(\tau\Rightarrow\tau\Rightarrow\tau)\Rightarrow(\tau\Rightarrow\tau\Rightarrow\tau)$,
for some basic type $\tau$.

A straightforward extension like this excludes from the calculus several valid
interesting terms. For example, let $t$ be a term typed with $\forall
X.A\Rightarrow B$ for some $A$. It could be applied not only to terms typed
with $[C/X]A$, but also to a linear combination $\sum_i\alpha_i.r_i$ as soon as
each $r_i$ is typed with some $[C_i/X]A$. Indeed, $t \sum_i\alpha_i.r_i$
reduces to $\sum_i\alpha_i.tr_i$. However, $\sum_i\alpha_i.r_i$ may have type
$\forall X.A$, but from there, only one $C_i$ can be chosen to type the whole
term. A solution to this lack of expressivity is provided by the Vectorial
calculus~\cite{ArrighiDiazcaroValironIC17}, which types this term with
$\sum_i\alpha_i.[C_i/X]A$.

The Vectorial calculus introduces linear combinations of types, in the same way
as Lineal considers linear combinations of terms. However, due to the
application group of rewrite rules, term variables must be typed only with
types that are not linear combinations of types (here called ``unit types'').
Indeed, suppose we admit variables of type $\alpha.U$, so $\lambda x.x+y$ is
typed with $(\alpha.U)\Rightarrow (\alpha.U)+V$, where $V$ is the type of $y$.
Then, if $u$ is typed with $U$, we may expect $\alpha.u$ to be typed with
$\alpha.U$, and so $(\lambda x.x+y)(\alpha.u)$ has type $(\alpha.U)+V$. However, 
\[
  (\lambda x.x+y)(\alpha.u)\lra\alpha.(\lambda x.x+y)u\lra \alpha.(u+y)\lra\alpha.u+\alpha.y
\]
so its type should be $\alpha.U+\alpha.V$ instead.

Indeed, the abstracted variable $x$ has been substituted by $u$ and not by
$\alpha.u$ during reduction, and so, the type of the abstraction should reflect
this being just $U\Rightarrow U+V$. Therefore, term variables must be typed
with unit types.

On the other hand, type variables do not always need to be unit types. For
example, we may consider variables $\mathcal X$, $\mathcal Y$, which can be
substituted only by unit types, and variables $\mathbb X$, $\mathbb Y$, which
can be substituted by any type. 

For instance, the Hadamard term $H$ from Equation~\eqref{eq:H} may be typed as follows.  Let 
$\mathcal T=\forall \mathcal X.\forall\mathcal Y.\mathcal X\Rightarrow \mathcal Y\Rightarrow \mathcal X$ and 
$\mathcal F=\forall \mathcal X.\forall\mathcal Y.\mathcal X\Rightarrow \mathcal Y\Rightarrow \mathcal Y$. 
Also, if $t$ has type $A$, we let thunks $[t]$ be typed with $[A]$, which is
just a notation for $(\forall\mathcal X.\mathcal X\Rightarrow\mathcal
X)\Rightarrow A$.  Then, 
\[
  \vdash H : \forall\mathbb X.\left(\left[\frac 1{\sqrt 2}.\mathcal T+\frac 1{\sqrt 2}.\mathcal F\right]\Rightarrow\left[\frac 1{\sqrt 2}.\mathcal T+\frac {-1}{\sqrt 2}.\mathcal F\right]\Rightarrow\left[\mathbb X\right]\right)\Rightarrow\mathbb X
\]

The full grammar of types is the following.
\begin{align*}
  A &:= U\mid\alpha.A\mid A+A\mid \mathbb X &\textrm{Types}\\
  U &:= \mathcal X\mid U\Rightarrow A\mid\forall\mathcal X.U\mid\forall\mathbb X.U & \textrm{Unit types}
\end{align*}
where we use $A,B,C$ for types, $U,V,W$ for unit types, $\mathbb X,\mathbb
Y,\mathbb Z$ for variables, and $\mathcal X,\mathcal Y,\mathcal Z$ for unit
variables. We use $X,Y,Z$ to refer to both kinds of variables.

There is also an equivalence between types given by the axioms of vector spaces
as follows.
\[
  \begin{array}{r@{\ }l@{\qquad\qquad}r@{\ }l@{\qquad\qquad}r@{\ }l}
    1.A &\equiv A 				&\alpha.A+\beta.B &\equiv \alpha.(A+B) 		& A+B &\equiv B+A\\
    \alpha.(\beta.A) &\equiv (\alpha\beta).A    & \alpha.A+\beta.A &\equiv (\alpha+\beta).A	& A+(B+S) &\equiv (A+B)+S
  \end{array}
\]
There is no general null type $0$, but one type $0$ for each type $A$. So, the term
$\vec 0$ is typed with the rule
\[
  \infer{\Gamma\vdash\vec 0:0.A}{\Gamma\vdash t:A}
\]

Taking again the example that started this section, to type
$t\sum_i\alpha_i.r_i$ where $\vdash t:\forall X.A\Rightarrow B$, and for each
$i$,  $\vdash r_i:[C_i/X]A$ we consider an $\Rightarrow$-elimination typing
rule such as
\[
  \infer{\Gamma\vdash tr:\sum_i\alpha_i.[C_i/X]B}
  {
    \Gamma\vdash t:\forall X.A\Rightarrow B
    &
    \Gamma\vdash r:\sum_i\alpha_i.[C_i/X]A
  }
\]
which is not only an $\Rightarrow$-elimination but also a $\forall$-elimination
at the same time.

Next, we also consider that $t$ can be a linear combination of terms, and,
since it is also a $\forall$-elimination, we may have more variables to
replace, which gives us the more general rule
\[
  \infer{\Gamma\vdash tr:\sum_i\sum_j\alpha_i\beta_j.[\vec C_j/\vec X]B_i}
  {
    \Gamma\vdash t:\sum_i\alpha_i.\forall\vec X.A\Rightarrow B_i
    &
    \Gamma\vdash r:\sum_j\alpha_j.[\vec C_j/\vec X]A
  }
\]
where, as usual, notation $\forall\vec X.A$ stands for $\forall X_1.\dots.\forall X_n.A$
and $[\vec C/\vec X]$ for $[C_1/X_1]\cdots[C_n/X_n]$.

The main theorem in~\cite{ArrighiDiazcaroValironIC17} is a strong normalisation
result for this calculus. However, subject reduction is proved only to some
extent. Indeed, if $t$ has both types $A$ and $B$, $\alpha.t+\beta.t$ may be
typed with $\alpha.A+\beta.B$, while its reduct, $(\alpha+\beta).t$ cannot. So,
a weaker subject reduction is proved, which states that if $\Gamma\vdash t:A$
and $t\lra r$, then $\Gamma\vdash r:B$ with $B$ related to $A$ by an ad-hoc
relation. Later, \cite{NoriegaDiazcaro20} slightly modifies Vectorial obtaining
a proper subject reduction result, without losing its main properties.

Some fragments of Vectorial are the Scalar type
system~\cite{ArrighiDiazcaroLMCS12}, which only includes scalars on types but
not sums. The Scalar type system can track the ``weight'' of a type, defined as
the $\ell_1$-norm of a term, which is useful, for example, to define a
probabilistic type system out of Lineal, by enforcing $\ell_1$-norm to be equal
to $1$ on each term. Another fragment is the Additive type
system~\cite{DiazcaroPetitWoLLIC12}, with sums but not scalars, which is shown
to be equivalent to System F with pairs.

\section{Lambda-\texorpdfstring{$\mathcal S$}{S}: Lineal plus measurement}
\label{sec:LambdaS}

Lineal takes care of three of the four postulates stated in
Section~\ref{sec:QC}. 
\begin{itemize}
  \item Postulate~\ref{pos:one}, referring to state vectors, can be encoded in
    several ways in Lineal. For example, we can take $\ket 0:=\lambda x.\lambda
    y.x$ and $\ket 1:=\lambda x.\lambda y.y$, and then make its linear
    combination to express any vector in $\mathbb C^2$. Also, for $\vec
    v\in\mathbb C^{2^n}$ it suffices to consider the basis $\{\lambda\vec
    x.x_1,\dots,\lambda\vec x.x_{2^n}\}$ where $\lambda\vec x.x_i$ stands for
    $\lambda x_1.\dots.\lambda x_{2^n}.x_i$.
  \item Postulate~\ref{pos:two}, referring to unitary matrices, can be encoded
    easily also. For example, any $U=(\alpha_{ij})_{ij}\in\mathbb
    C^2\times\mathbb C^2$, can be written as
    \[
      U := \lambda x.\left\{x \left[\alpha_{00}.\ket 0 + \alpha_{01}.\ket 1\right] \left[\alpha_{10}.\ket 0 + \alpha_{11}.\ket 1\right]\right\}
    \]
  \item Postulate~\ref{pos:four}, referring to composing systems, can be also
    represented by taking the usual Church encoding for pairs, since the
    linearity given by the application group of rewrite rules will make these
    pairs bi-linear (i.e.~left and right linear), as a tensor product.
\end{itemize}
The only missing postulate is Postulate~\ref{pos:three}, referring to
measurements. Suppose we want to add a term $\pi$ representing a measurement
operator in the computational basis such that $\pi (\alpha.\ket 0+\beta.\ket
1)$ reduces to $\ket 0$ with probability $|\alpha|^2$ and to $\ket 1$ with
probability $|\beta|^2$ (if $|\alpha|^2+|\beta|^2=1$, otherwise it suffices to
divide this term by $|\alpha|^2+|\beta|^2$ before reducing). The problem is
that $\pi$ should not be linear, but then, $\lambda x.\pi x$ would not behave
as $\pi$ since
\[
  (\lambda x.\pi x)(\alpha.\ket 0+\beta.\ket 1)
  \lra^*
  \alpha.(\lambda x.\pi x)\ket 0+\beta.(\lambda x.\pi x)\ket 1
  \lra^*
  \alpha.\pi\ket 0+\beta.\pi\ket 1
  \lra^*
  \alpha.\ket 0+\beta.\ket 1
\]
which is definitely not what we meant to do.  Indeed, $(\lambda x.\pi x)r$
should reduce to $\pi r$ whatever $r$ is, even if it is a superposition.  This
jeopardises the entire encoding. As we pointed out in
Remark~\ref{rmk:no-cloningLineal}, the algebraic linearity (i.e.~the
application group of rewrite rules) is needed to forbid cloning. An alternative
solution, first described in Lambda-$\mathcal
S$~\cite{DiazcaroDowekTPNC17,DiazcaroDowekRinaldiBIO19}, is to allow for
certain functions to be call-by-name, that is, $(\lambda x.t)r$ reduces to
$(r/x)t$ wherever $r$ is, at the condition that $x$ appears at most once in
$t$, which also forbids duplication. With this goal in mind, Lambda-$\mathcal
S$ is typed with simple types, but adding a new symbol $S$ which marks the
``superpositions'', as those types that forbid duplication. The grammar of
types is given by
\begin{align*}
  \Psi &:=\mathbb B\mid \Psi\times\Psi\mid S\Psi & \textrm{Qubit types}\\
  A&:=\Psi\mid\Psi\Rightarrow A\mid A\times A\mid SA &\textrm{Types}
\end{align*}
Since superposition of superpositions is still a superposition, $SSA = SA$, and
since a basis term such as $\ket 0$ can be seen as the superposition of $1.\ket
0+0.\ket 1$, $A\leq SA$.

Lambda-$\mathcal S$ includes constants $\ket 0$ and $\ket 1$, instead of using
Church encodings, as well as an if-then-else construction. However, the most
interesting characteristic, other than the application group from Lineal, is
the fact that there are two beta-reductions, depending on types. If $b$ is a
basis term, then $(\lambda x:\mathbb B^n.t)b$ reduces to $(b/x)t$, which is
exactly Lineal's beta-reduction. Instead, $(\lambda x:S\Psi.t)r$ reduces
directly to $(r/x)t$, for any $r$. However, the typing system ensures that $x$
does not appear more than once in $t$ in this second case.  This way,
\[
  (\lambda x:\mathbb B.t)(\alpha.\ket 0+\beta.\ket 1)
  \lra
  \alpha.(\lambda x:\mathbb B.t)\ket 0+\beta.(\lambda x:\mathbb B.t)\ket 1
  \lra^*\alpha.(\ket 0/x)t+\beta.(\ket 1/x)t
\]
while
\[
  (\lambda x:S\mathbb B.\pi x)(\alpha.\ket 0+\beta.\ket 1)\lra\pi(\alpha.\ket 0+\beta.\ket 1)
\]
as expected.

The term $H$ from Example~\ref{ex:H} is still valid, typed with $\mathbb
B\Rightarrow S\mathbb B$ on this system.

A first denotational semantics (in environment style) is given where the type
$\mathbb B$ is interpreted as $\sem{\mathbb B}=\{\ket 0,\ket 1\}$ while $SA$ is
interpreted as $\mathsf{Span}\sem A$, the vector space generated by $\sem A$.
For example, $\sem{S\mathbb B}=\mathbb C^2$.
In~\cite{DiazcaroMalherbeLSFA18,DiazcaroMalherbe2021} a concrete categorical
interpretation is given, where $S$ is considered as a function of an adjunction
between the category $\mathbf{Set}$ and the category $\mathbf{Vec}$.
Explicitly, when we evaluate $S$ we obtain formal linear combinations of
elements of a set with complex numbers as coefficients. The other functor in
the adjunction, $U$ is just the forgetful functor allowing us to forget its
vectorial structure.

Later, in~\cite{DiazcaroMalherbeACS2020}, an abstract categorical semantics of
Lambda-$\mathcal S$ has been defined. The main structural feature of such a
model is that it is expressive enough to describe the bridge between the
property-less elements such as $\alpha.v+\beta.v$, without any equational
theory, and the result of its algebraic manipulation into $(\alpha+\beta).v$,
explicitly controlling its interaction. 

A distinctive design choice of Lambda-$\mathcal S$ is that we mark the
superpositions, which are the non-duplicable elements. This is somehow the
opposite of what is done in Linear Logic, where a bang $!$ marks the duplicable
elements.  In fact, it is common that intuitionistic linear models (linear as
in linear-logic) are obtained by a monoidal comonad determined by a monoidal
adjunction $(S,m)\dashv (U,n)$, that is, the bang $!$ is interpreted by the
comonad $SU$ (see for example~\cite{BentonCSL94}). Instead, a crucial point
of our model of Lambda-$\mathcal S$ is to consider the monad $US$ for the
interpretation of $S$, determined by a similar monoidal adjunction. This
implies that on the one hand we have tight control of the Cartesian structure
of the model (i.e.~duplication, etc.) and on the other hand superpositions live
in some sense inside the classical world determined externally by classical
rules until we decide to explore it. This is given by the following composition
of maps:
\[
  US\mathbb B\times US\mathbb B\xrightarrow{n} U(S\mathbb B\otimes S\mathbb B)\xrightarrow{Um} US(\mathbb B\times\mathbb B)
\]

\begin{remark}
  \label{rmk:normOne}
  Lambda-$\mathcal S$ includes the four postulates, plus classical computation,
  since simply typed lambda calculus is a subset of it. However, the first
  postulate is included in a too general way, as with Lineal: any superposition
  $\alpha.\ket 0+\beta.\ket 1$ is valid, for any $\alpha$ and $\beta$, so,
  instead of taking norm-$1$ vectors (cf.~Postulate~\ref{pos:one}), we are
  taking any vectors. Moreover, an abstraction $\vdash\lambda x:\mathbb
  B.t:\mathbb B\Rightarrow S\mathbb B$ may not represent a unitary matrix. For
  example, $\lambda x:\mathbb B.\mathsf{if}\ x\ \mathsf{then}\ \ket\psi\
  \mathsf{else}\ \ket\varphi$ is typable, but does not represent a unitary map
  unless $\ket\psi\perp\ket\varphi$. To ensure unitarity we should add some
  ad-hoc restrictions to ensure that the branches of an if-then-else are
  orthogonal.  However, the orthogonality of two values is not so hard to
  define, but the orthogonality between arbitrary programs does not seem to be
  an easy task.

  QML~\cite{AltenkirchGrattageLICS05} introduced some syntactic judgements of
  the form $t\perp r$ for a limited subset of terms. In the next section we see
  how to produce a calculus from a model where only norm-$1$ vectors are
  allowed.
\end{remark}

\section{Realisability to the rescue}
\label{sec:LambdaSone}

From Remark~\ref{rmk:normOne} it is clear that we need a quite non-standard
restriction to the linear combinations. The fact that not every vector in
$\mathbb C^2$ is a qubit, but only those in the unitary sphere, is a hard
condition to ask in a type system. For example, in Vectorial (see
Section~\ref{sec:Vectorial}) we may restrict types $\alpha.A+\beta.B$ to the
case $|\alpha|^2+|\beta|^2=1$, but if $A=B=\forall\mathcal X.\mathcal
X\Rightarrow\mathcal X\Rightarrow\mathcal X$, this condition is not enough, for
example, this is a valid type for $\alpha.\ket 0+\beta.\ket 0\lra
(\alpha+\beta).\ket 0$, which does not have norm $1$ if
$|\alpha|^2+|\beta|^2=1$.

The restriction of Lambda-$\mathcal S$ to norm-$1$ vectors has been obtained
in~\cite{DiazcaroGuillermoMiquelValironLICS19} by means of realisability
techniques~\cite{KleeneJSL45,Vonoosten08,KrivinePS09,MiquelTLCA11}, which
proved to be a great method to add any kind of ad-hoc restrictions in a clean
and easy way. The technique can be summarised as follows: 
\begin{enumerate}
  \item Take an untyped machine (i.e.~a calculus with a fixed strategy). It is
    {\em confluent by definition}.
  \item Then, define a grammar of types and give an interpretation $\sem A$ for
    each type $A$ as a set of values, for example, allow only for values $v$
    such that $\|v\|=1$. So, it has {\em norm-$1$ by definition}.
  \item Then, define that a term $t$ is a realizer of $A$ (notation $t\vDash
    A$) whenever $t\lra^* v\in\sem A$. A sequent $\Gamma\vdash t:A$ is defined
    to be valid if for any valid substitution $\sigma$ in $\Gamma$, $\sigma
    t\vDash A$.  So, typed terms are {\em strongly normalising by definition},
    and {\em subject reduction is also ensured by definition}.
  \item\label{it:stepfour} Finally, a typing rule of the kind
    \[
      \infer{\Gamma\vdash t:A}{\Delta\vdash r:B}
    \]
    is valid as soon as $\Delta\vdash r:B$ implies $\Gamma\vdash t:A$. This
    way, typing rules become {\em theorems} (potentially infinite many of
    them). 
\end{enumerate}

So, instead of defining typing rules and proving all its desirable properties,
we give the desirable properties by definition and prove the typing rules.  Of
course this recipe does not tell us how to statically determine whether $t\perp
r$ for some arbitrary terms $t$ and $r$, but it gives us a way to check whether
any given typing rule is valid, and forces the system to only let pass whatever
terms we want (for example, only allow for terms of type $S\mathbb B\Rightarrow
S\mathbb B$ if those terms represent unitary maps).

The notation in this system is a bit modified from that of Lambda-$\mathcal S$,
so instead of $SA$ we write $\sharp A$ for the set $\mathsf{Span}\sem
A\cap\{v\mid\|v\|=1\}$ and we even have a type $\flat A$ which is interpreted
as the smallest set $V$ of values such that $\sharp V$ contains $\sem A$. We do
not give more details in this quick overview, since there are many, but it is
worth mentioning that a quantum lambda calculus in the spirit
of the QLC~\cite{SelingerValironMSCS06} is translated into this calculus
in~\cite{DiazcaroGuillermoMiquelValironLICS19}.

As mentioned in the step~\ref{it:stepfour} of the above enumeration, the typing
rules are potentially infinite. So, in~\cite{DiazcaroMalherbe2020b} we
extracted a finite fixed type system, defining Lambda-$\mathcal S_1$, and gave
a categorical interpretation for it.  The model developed has some common
grounds with the concrete model of Lambda-$\mathcal
S$~\cite{DiazcaroMalherbeLSFA18,DiazcaroMalherbe2021}, however, the chosen
categories this time are not $\mathbf{Set}$ and $\mathbf{Vec}$, but categories
that use the fact that values in this calculus form a ``distributive-action
space'' (an algebraic structure similar to a vector space, where its additive
structure is a semigroup). The two categories in the constructed adjunction
are defined in terms of $\vec{\mathsf V}$, the set of values in the calculus,
and their linear combinations (such a set forms a distributive-action space).
Then, the categories for the adjunction are defined by:
\begin{itemize}
  \item $\mathbf{Set}_{\vec{\mathsf V}}$: a category whose objects are the
    non-empty parts of $\vec{\mathsf V}$, and whose arrows are the arrows in
    $\mathbf{Set}$ that can be defined in Lambda-$\mathcal S_1$.  This category
    also includes a product $\boxtimes$, which is the set of separable tensor
    products. 
  \item $\mathbf{SVec}_{\vec{\mathsf V}}$: a category whose objects are the
    sub-distributive action spaces of $\vec{\mathsf V}$, and whose arrows are
    the linear maps which can be defined in Lambda-$\mathcal S_1$. It also
    includes a tensor product $\otimes$, which can also be defined as the span
    of the product $\boxtimes$.
\end{itemize}

The main novelty and contribution of~\cite{DiazcaroMalherbe2020b} is presenting
a model for quantum computing in the quantum control paradigm, which is show to
be complete on qubits in the sense that if two closed terms with qubit types
are interpreted by the same arrows in the model, then those terms are
computationally equivalent.

\section{Sup: A new connective in Natural Deduction}
\label{sec:Sup}
As mentioned in the introduction, one of the main goals of the quantum control
approach to the lambda calculus is to envisage a quantum computational logic
(i.e.~a quantum logic founded by quantum computing and an extension to the
Curry-Howard correspondence). In Section~\ref{sec:Vectorial} we showed
Vectorial, where if $A$ and $B$ are two propositions, so is $\alpha.A+\beta.B$.
So, we could restrict the valid propositions $A$ to those for whom there are
valid proof terms $t$ such that $\vdash t:A$ and $t$ represents a quantum
state. Vectorial can be seen as the propositional logic of vector spaces.  In
Section~\ref{sec:LambdaS} we gave another form to the superposition of
propositions: if $A$ is a proposition, then $SA$ is a superposition of
propositions $A$. In this case the superposition symbol is unary, and from
several proofs $t_i$ of $A$ we can construct a proof $\sum_i\alpha_i.t_i$ of
$SA$. Similarly, in Section~\ref{sec:LambdaSone} we not only gave the
superpositions $SA$ (now noted $\sharp A$), but also another unary symbol
$\flat$, from where if $A$ is a proposition, $\flat A$ is the proposition whose
set of proofs is the minimum set such that the unary span of such a set is the
set of proofs of $\sharp A$.

In all these previous works we started from the lambda calculus Lineal, and
worked out a logic (a type system) from it. In~\cite{DiazcaroDowekICTAC2021} we
started from the other end. Starting from Natural Deduction we worked out a new
connective for superpositions, and from there we gave a proof system which can
be used to encode quantum computing on it.

The rest of this section paraphrases the extended abstract
of~\cite{DiazcaroDowekICTAC2021} we have presented at QPL 2021. Two nice video
presentations given by Gilles Dowek can be found
at~\cite{DowekQPL2021,DowekICTAC2021}.

\paragraph*{Insufficient, harmonious, and excessive connectives}
In natural deduction, to prove a proposition $C$, the elimination rule of a
connective $\vartriangle$ requires a proof of $A \vartriangle B$ and a proof of
$C$ using, as extra hypotheses, exactly the premises needed to prove the
proposition $A \vartriangle B$, with the introduction rules of the connective
$\vartriangle$. This principle of inversion, or of harmony, has been introduced
by Gentzen \cite{Gentzen} and developed, among others, by Prawitz
\cite{Prawitz} and Dummett \cite{Dummett} in natural deduction, by Miller and
Pimentel \cite{MillerPimentel} in sequent calculus, and by Read
\cite{Read04,Read10,Read} for the rules of equality.

For example, to prove the proposition $A \wedge B$, the introduction rule of
the conjunction requires a proof of $A$ and a proof of $B$, hence, to prove a
proposition $C$, the generalised elimination rule of the conjunction
\cite{SchroederHeister,Parigot,NegriPlato} requires, a proof of $A\wedge B$ and
a proof of $C$, using, as extra hypotheses, the propositions $A$ and $B$
\[
  \irule{\Gamma \vdash A \wedge B & \Gamma, A, B \vdash C}
  {\Gamma \vdash C}
  {\mbox{$\wedge$-e}}
\]
This principle of inversion permits to define a cut elimination process where
the proof
\[
  \irule{\irule{\irule{\pi_1}{\Gamma \vdash A}{}
      &
      \irule{\pi_2}{\Gamma \vdash B}{}
    }
    {\Gamma \vdash A \wedge B}
    {\mbox{$\wedge$-i}}
    &
    \irule{\pi_3}{\Gamma, A, B \vdash C}{}
  }
  {\Gamma \vdash C}
  {\mbox{$\wedge$-e}}
\]
reduces to $(\pi_1/A,\pi_2/B)\pi_3$.

In the same way, to prove the proposition $A \vee B$, the introduction rules of
the disjunction require a proof of $A$ or a proof of $B$, hence, to prove a
proposition $C$, the elimination rule of the disjunction requires a proof of
$A\vee B$ and two proofs of $C$, one using, as extra hypothesis, the
proposition $A$ and the other the proposition $B$
\[
  \irule{\Gamma \vdash A}
  {\Gamma \vdash A \vee B}
  {\mbox{$\vee$-i1}}
  \qquad
  \qquad
  \irule{\Gamma \vdash B}
  {\Gamma \vdash A \vee B}
  {\mbox{$\vee$-i2}}
  \qquad
  \qquad
  \irule{\Gamma \vdash A \vee B & \Gamma, A \vdash C & \Gamma, B \vdash C}
  {\Gamma \vdash C}
  {\mbox{$\vee$-e}}
\]
and a cut elimination process can be defined similarly.

We also can imagine connectives that do not verify this inversion principle,
because the introduction rules require an insufficient amount of information
with respect to what the elimination rule provides, as extra hypotheses, in the
required proof of $C$. An example of such an {\em insufficient} connective is
Prior's {\it tonk} \cite{Prior}, with the introduction and elimination rules as
follows 
\[
  \irule{\Gamma \vdash A}
  {\Gamma \vdash A~\mbox{\it tonk}~B}
  {\mbox{{\it tonk}-i}}
  \qquad
  \qquad
  \irule{\Gamma \vdash A~\mbox{\it tonk}~B & \Gamma, B \vdash C}
  {\Gamma \vdash C}
  {\mbox{{\it tonk}-e}}
\]
where the elimination rule requires a proof of $A~\mbox{\it tonk}~B$ and a
proof of $C$, using the extra hypothesis $B$, that is not required in the proof
of $A~\mbox{\it tonk}~B$, with the introduction rule. For such connectives,
cuts {\it tonk}-i / {\it tonk}-e cannot be reduced.

But, it is also possible that a connective does not verify the inversion
principle because the introduction rules require an excessive amount of
information.  An example of such an {\em excessive} connective is the
connective $\odot$ that has the introduction rule of the conjunction and the
elimination rule of the disjunction
\[
  \irule{\Gamma \vdash A & \Gamma \vdash B}
  {\Gamma \vdash A \odot B}
  {\mbox{$\odot$-i}}
  \qquad\qquad
  \irule{\Gamma \vdash A \odot B & \Gamma, A \vdash C & \Gamma, B \vdash C}
  {\Gamma \vdash C}
  {\mbox{$\odot$-e}}
\]
In this case, cuts can be eliminated. Moreover, several cut elimination
processes can be defined, exploiting, in different ways, the excess of the
connective. For example, the $\odot$-cut
\[
  \irule{\irule{\irule{\pi_1}{\Gamma \vdash A}{}
      &
      \irule{\pi_2}{\Gamma \vdash B}{}
    }
    {\Gamma \vdash A \odot B}
    {\mbox{$\odot$-i}}
    &
    \irule{\pi_3}
    {\Gamma, A \vdash C}{} & \irule{\pi_4}{\Gamma, B \vdash C}
    {}
  }
  {\Gamma \vdash C}
  {\mbox{$\odot$-e}}
\]
can be reduced to $(\pi_1/A)\pi_3$, it can be reduced to $(\pi_2/A)\pi_4$, it
also can be reduced, non-deterministically, either to $(\pi_1/A)\pi_3$ or to
$(\pi_2/A)\pi_4$. Finally, to keep both proofs, we can add a structural rule
\[
  \vcenter{\irule{\Gamma \vdash A & \Gamma \vdash A}
    {\Gamma \vdash A}
  {\mbox{parallel}}}
  \qquad\qquad\textrm{and reduce it to}\qquad\qquad
  \vcenter{\irule{\irule{(\pi_1/A)\pi_3}
      {\Gamma \vdash C}
      {}
      & 
      \irule{(\pi_2/B)\pi_4}
      {\Gamma \vdash C}
      {}
    }
    {\Gamma \vdash C}
  {\mbox{parallel}}}
\]

\paragraph*{Information loss}
\label{informationloss}
With harmonious connectives, when a proof is built with an introduction rule,
the information contained in the proofs of the premises of this rule is
preserved. For example, the information contained in the proof $\pi_1$ is {\em
present} in the proof $\pi$
\[
  \irule{\irule{\pi_1}{\Gamma \vdash A}{}
    &
    \irule{\pi_2}{\Gamma \vdash B}{}
  }
  {\Gamma \vdash A \wedge B}
  {\mbox{$\wedge$-i}}
\]
in the sense that $\pi_1$ is a subtree of $\pi$. But it is moreover {\em
accessible}, in the sense that, for all $\pi_1$, putting the proof $\pi$ in the
right context yields a proof that reduces to $\pi_1$.  And the same holds for
the proof $\pi_2$.

The situation is different with an excessive connective: the excess of
information, required by the introduction rule, and not returned by the
elimination rule in the form of an extra hypothesis, in the required proof of
$C$, is lost.  For example, the information contained in the proofs $\pi_1$ and
$\pi_2$ is present in the proof
\[
  \irule{\irule{\pi_1}{\Gamma \vdash A}{}
    & 
    \irule{\pi_2}{\Gamma \vdash B}{}
  }
  {\Gamma \vdash A \odot B}
  {\mbox{$\odot$-i}}
\]
but its accessibility depends on the way we decide to reduce the cut 
\[
  \irule{\irule{\irule{\pi_1}{\Gamma \vdash A}{}
      &
      \irule{\pi_2}{\Gamma \vdash B}{}
    }
    {\Gamma \vdash A \odot B}
    {\mbox{$\odot$-i}}
    &
    \irule{\pi_3}
    {\Gamma, A \vdash C}{} & \irule{\pi_4}{\Gamma, B \vdash C}
    {}
  }
  {\Gamma \vdash C}
  {\mbox{$\odot$-e}}
\]
If we reduce it systematically to $(\pi_1/A)\pi_3$, then the information
contained in $\pi_1$ is accessible, but that contained in $\pi_2$ is not.  If
we reduce it systematically to $(\pi_2/A)\pi_4$, then the information contained
in $\pi_2$ is accessible, but not that contained in $\pi_1$.  If we reduce it
not deterministically to $(\pi_1/A)\pi_3$ or to $(\pi_2/A)\pi_4$, then the
information contained in both $\pi_1$ and $\pi_2$ is accessible but 
non-deterministically. If we reduce it with parallel, then the information
contained in both $\pi_1$ and $\pi_2$ is inaccessible.

So, while harmonious connectives, that verify the inversion principle, model
information preservation, reversibility, and determinism, these excessive
connectives, that do not verify the inversion principle, model information
erasure, non-reversibility, and non-determinism.  Such information erasure,
non-reversibility, and non-determinism, occur, for example, in quantum physics,
where the measurement of the superposition of two states does not yield both
states back.

\paragraph*{A quantum language with $\odot$} As we have seen in the
previous sections, several programming languages have been designed to express
quantum algorithms with quantum control.  In particular, in Lambda~$\mathcal S$
(see Section~\ref{sec:LambdaS}), the measurement operator $\pi$ comes together
with the rule reducing $\pi(t+r)$ non-deterministically to $t$ or to $r$.

The superposition $t + r$ can be considered as the pair $\pair{t}{r}$, as
stated by~\cite{DiazcaroPetitWoLLIC12}.  Hence, it should have the type $A
\wedge A$. In other words, it is a proof-term of the proposition $A \wedge A$.
In System I (first introduced in~\cite{SystemI} and later extended
in~\cite{SystemIeta,SotilleDiazcaroMartinezlopezIFL20,DiazcaroMartinezlopezIFL15}),
various type-isomorphisms have been introduced, in particular the commutativity
isomorphism $A \wedge B \equiv B \wedge A$, hence $t + r \equiv r + t$. In such
a system, where $A \wedge B$ and $B \wedge A$ are identical, it is not possible
to define the two elimination rules, as the two usual projections rules $\pi_1$
and $\pi_2$ of the $\lambda$-calculus. They were replaced with a single
projection parametrised with a proposition $A$: $\pi_A$, such that if $t$ is
typed by $A$ and $r$ by $B$ then $\pi_A(t + r)$ reduces to $t$ and $\pi_B(t +
r)$ to $r$.  When $A = B$, so $t$ and $r$ both have type $A$, the proof-term
$\pi_A(t + r)$ reduces, non-deterministically, to $t$ or to $r$.  Thus, this
modified elimination rule behaves like a measurement operator.

These works on Lambda-${\mathcal S}$ and System I brought to light the fact
that the pair superposition / measurement, in a quantum programming language,
behaves like a pair introduction / elimination, for some connective, in a proof
language, as the succession of a superposition and a measurement yields a term
that can be reduced.  In System I, the assumption was made that this connective
was a commutative conjunction, with a modified elimination rule, leading to a
non-deterministic reduction.

However, as the measurement of the superposition of two states does not yield
both states back, this connective should probably be excessive. Moreover, as,
to build the superposition $a . \ket 0 + b . \ket 1$, we need both $\ket 0$ and
$\ket 1$ and the measurement, in the basis $\ket 0, \ket 1$, yields either
$\ket 0$ or $\ket 1$, this connective should have the introduction rule of the
conjunction, and the elimination rule of the disjunction, that is that it
should be the connective $\odot$.

In~\cite{DiazcaroDowekICTAC2021}, we present a propositional logic with the
connective $\odot$, a language of proof-terms, the $\odot$-calculus (read:
``the sup-calculus''), for this logic, and we prove a cut elimination theorem.
We then extend this calculus into the $\odot^{\mathbb C}$-calculus
(sup-$\mathbb C$-calculus), introducing scalars to quantify the propensity of a
proof to reduce to another and show that its proof language forms the core of a
quantum programming language. 

In particular, $\vdash\alpha.*:\top$ making of $\top$ a representation of
$\mathbb C$. In addition, while $\top\vee\top$ represent the booleans in the
lambda calculus, $\top\odot\top$ represents the qubits, $\mathbb C^2$. We use
$+$ to the proof term of the propositions $\odot$, and so
$\vdash:\alpha.*+\beta.*:\top\odot\top$ represents the vector $\alpha.\ket
0+\beta.\ket 1$.

The elimination of $\odot$ has two proof terms:
\[
  \vcenter{
    \infer{\Gamma\vdash\delta_{\odot}^\parallel(t,[x]r,[y]s):C}
    {
      \Gamma\vdash t:A\odot B
      &
      \Gamma,x:A\vdash r:C
      &
      \Gamma,y:A\vdash r:C
    }
  }
  \qquad
  \vcenter{
    \infer{\Gamma\vdash\delta_{\odot}(t,[x]r,[y]s):C}
    {
      \Gamma\vdash t:A\odot B
      &
      \Gamma,x:A\vdash r:C
      &
      \Gamma,y:A\vdash r:C
    }
  }
\]
The first one reduces with the following rule
\[
  \delta_\odot^\parallel(\alpha.t+\beta.r,[x]s_1,[y]s_2) \lra \alpha.(t/x)s_1\parallel \beta.(r/x)s_2
\]
where the $\parallel$ is such that it behaves as a vectorial sum. 

The second one reduces with the following rules instead
\begin{align*}
  \delta_\odot(\alpha.t+\beta.r,[x]s_1,[y]s_2) &\lra (t/x)s_1\\
  \delta_\odot(\alpha.t+\beta.r,[x]s_1,[y]s_2) &\lra (r/x)s_2
\end{align*}
where the first reduction happens with probability
$\frac{|\alpha|^2}{|\alpha|^2+|\beta|^2}$ and the second with probability
$\frac{|\beta|^2}{|\alpha|^2+|\beta|^2}$.

The term $\delta_\odot^\parallel$ is then used to encode quantum gates and
$\delta_\odot$ to encode the measurement.  See~\cite{DiazcaroDowekICTAC2021}
for more details.

\section{Towards quantum recursive types}
\label{sec:recursive}
A while language called qGCL with quantum control has been introduced
in~\cite{SandersZulianiMPC00}, and an entire book on the subject has been
written by Ying~\cite{FOQ}. The idea is to consider an infinite-dimensional
Hilbert space and the while is guarded by a binary measurement on one qubit,
which stops when the outcome is $0$.

In~\cite{SabryValironVizzottoF18}, a similar idea of quantum control loop has
been taken to the lambda calculus, by endowing a typed, reversible, algebraic
lambda calculus of structural recursive fixpoints linked to the convergence of
sequences in infinite-dimensional Hilbert spaces. A categorical semantics of
this language is given in~\cite{LemonnierChardonnetValiron21}.
In~\cite{ChardonnetSaurinValironRC20}, a language, based on this reversible
language, is typed in $\mu$MALL (linear logic extended with least and greatest
fixed points) allowing inductive and coinductive statements. While the paper
\cite{ChardonnetSaurinValironRC20} is subtitled ``work-in-progress'', it is a
firm first step towards quantum recursive types.

\section{Some final thoughts} \label{sec:Conclusion} This quick overview
aims to give a bird's eye view on the quest for a quantum computational logic.
There are several clues on how this logic should be. Either by representing
superposition of propositions as $\alpha.A+\beta.B$ (see
Section~\ref{sec:Vectorial}), as $SA$ or $\sharp A$ (see
Sections~\ref{sec:LambdaS} and \ref{sec:LambdaSone}) or by $A\odot B$ (see
Section~\ref{sec:Sup}), the many options present different approaches, but all
of them are founded by the Curry-Howard isomorphism: a proposition is a type
and a proof is a term.

There is still work to do. One may wonder if from $\odot$ we may define a
measurement in Vectorial, for example, where the $+$ symbol at
$\alpha.A+\beta.B$ can be eliminated by a disjunction elimination. Or what is
the meaning of $\odot$ in sequent calculus, what categorical construction may
model it, etc.

All of those are open problems we are willing to investigate.

\subsection*{Acknowledgements}
The author wishes to thank Eduardo Bonelli and Mauricio Ayala-Rinc\'on for the
inclusion of this overview in LSFA 2021 as an invited talk, and for their
useful comments.

\bibliographystyle{eptcs}
\bibliography{biblio}
\end{document}